# A Comprehensive Analysis of 2D&3D Video Watching of EEG Signals by Increasing PLSR and SVM Classification Results

*Corresponding Author:*
**Negin MANSHOURI**
Karadeniz Technical University
Faculty of Engineering
Department of Electrical and Electronics Engineering
61080 Trabzon/TURKEY
Phone (mobile): (+90) 5535791640
E-mail: n_manshoori90@yahoo.com

**Temel KAYIKCIOGLU**
Karadeniz Technical University
Faculty of Engineering
Department of Electrical and Electronics Engineering
61080 Trabzon/TURKEY
E-mail: tkayikci@ktu.edu.tr

# A Comprehensive Analysis of 2D&3D Video Watching of EEG Signals by Increasing PLSR and SVM Classification Results

Negin MANSHOURI, Temel KAYIKCIOGLU

*Abstract*

Despite the development of two and three dimensional (2D&3D) technology, it has attracted the attention of researchers in recent years. This research is done to reveal the detailed effects of 2D in comparison with 3D technology on the human brain waves. The impact of 2D&3D video watching using electroencephalography (EEG) brain signals is studied. A group of eight healthy volunteers with the average age of 31±3.06 years old participated in this three-stage test. EEG signal recording consisted of three stages: After a bit of relaxation (a), a 2D video was displayed (b), the recording of the signal continued for a short period of time as rest (c), and finally the trial ended. Exactly the same steps were repeated for the 3D video. Power spectrum density (PSD) based on short time Fourier transform (STFT) was used to analyze the brain signals of 2D&3D video viewers. After testing all the EEG frequency bands, delta and theta were extracted as the features. Partial least squares regression (PLSR) and Support vector machine (SVM) classification algorithms were considered in order to classify EEG signals obtained as the result of 2D&3D video watching. Successful classification results were obtained by selecting the correct combinations of effective channels representing the brain regions.

*1. Introduction*

As known, the brain is made up of billions of messengers that are called neurons. These cells are connected to other neurons by releasing the chemical signals; these measurable signals help assess the brain activity. Electroencephalography (EEG) is used to clarify how electrical sparks are between the brain neurons. By recording the electrical pattern of the neurons, this monitoring method informs us about the condition of the brain.

In spite of the weak spatial sensitivity of EEG, a great volume of research has been done using EEG due to its great potential, including portability, simplicity, high precision time measurements, etc. Also, the frequency bands of the EEG are classified into 5 categories: delta ($\delta$, 1-4 Hz), theta ($\theta$, 4-8 Hz), alpha ($\alpha$, 8-13 Hz), beta ($\beta$, 13-32 Hz) and gamma ($\gamma$, >32 Hz) [1], [2]. The usage of EEG in the brain-computer interface (BCI) technology shows the important role of these signals in the human life. This interface is a strong signal analyzer of the brain which, via translating its signals into commands of the output devices, can create miracles, especially in the lives of people with disabilities. An extensive field of application of EEG can be listed as follows: diagnostics [3], [4], [5], rehabilitation [6], cognitive training [7], biofeedback therapy [8], security [9] and neuro entertainment.

In addition to the medical applications of EEG, it can be used for neuro entertainment fields such as: neurogaming (brain-controlled game of Neurosky app), neuro toys (Puzzlebox, Star Wars-themed toys [10], virtual reality and art. Experiencing the virtual world has been made possible by using the VR headsets and EEG signals [11]. Also, these signals have contributed to the art industry, including music generation [12].

2D&3D technology analysis using brain signals has also attracted the attention of explorers. In order to understand this technology, scientific information is required on how the eye works. Stereovision (known as stereoscopic vision or stereopsis) is a normal human vision that has some amazing features. This kind of vision is a combination of identical images from the two eyes, with a slight angle change. In spite of great similarity between the views of two eyes, each eye captures exclusive visual information. This information is sent to the brain for processing. By matching up the similarities and adding in the small differences, a big difference is obtained in the final image. This image is the three-dimensional (3D) image. The relationship between this type of vision and the function of the brain has been the subject of many studies. Addition of the depth dimension perception is the reason why stereo vision is so rich and special. In terms of brain, stereo vision is a function that supports the cognitive processes of the brain [13].

EEG analysis of 2D&3D technology generally can be classified as eye fatigue detection of this technology, 2D & 3D game analysis, detection of power spectrum differences in brain waves and investigation of the effects of stereoscopic disparity on event-related potentials. A limited number of these studies have been focused on the comprehensive analysis of the brain signals in 2D&3D modes. In the game analysis [14], a team of researchers has proposed a new method by classifying 2D&3D data. In this research by using various nonlinear features, five physiological conditions are classified with a high percentage of accuracy. In [15], signal complexity was checked by two parameters (Hjorth-complexity and composite permutation entropy index). The complexity level has increased in the eye opened and 3D mode. In [16] by considering the power of neuronal oscillations and coherence, an increase in the power of the lower frequency at the frontal and occipital regions in 3D and the higher power of the upper frequency at temporal lobes for 2D were explored. Electric power of the brain waves as the result of watching the 2D, 2.5D and 3D motion pictures and also the comparison of these modes for alpha and gamma bands were done in [17]. The study of visual fatigue for a 2D display against 3D was performed in [18]. In terms of the educational aspect, brain signals recorded from the watching 2D&3D format were studied [19]. The support vector machine (SVM) classification results showed a slight difference in the 2D format compared with the 3D one. The vertical difference in 2D&3D (with three vertical disparities) images and its effect on visual discomfort was investigated in [20]. Due to the degree of depth in the 3D image, a significant increase in ERP components (P1), compared to 2D, was observed. By decreasing the perceptions of depth, the amplitude of this component can be modulated.

Looking at the screen from different distances and how it affected the human visual system were explored in [21]. In this study, watching time and screen distance were determined as important factors in 3D video watching and also it was shown that the effect of eye fatigue decreased when the screen distance increased in 3D.

The analysis of the signal response of the human brain in watching 2D&3D video is the main framework of this article. In fact, the main goal is to select the effective channels and EEG band to evaluate human cognitive and neural responses, with a reliance on the numerical results. In our previous study [22], the avatar video was displayed in the 2D&3D mode to the volunteers, the brain signals were then extracted by the fast Fourier transformation (FFT) feature and classified by "Classification Learner App". The classification accuracy in five EEG bands for channel T5 classified by Quadratic SVM was obtained as the highest. Kim and Lee [23] showed that in the case of watching 3D movie, the EEG power was increased compared to 2D. Also, these researchers illuminated that $\alpha$ band was less important than the $\beta$ band in terms of watching 2D&3D movies. Subasi [24] observed that the increase in the frequency of $\delta$ and $\theta$ was directly related to fatigue increase. Chen et al. [25] used four power indices of EEG to show that, when the participants were watching 3D films compared to 2D, the energy of $\alpha$ and $\beta$ frequency bands was greatly reduced and the $\delta$ band's energy increased significantly. However, the $\theta$ band's energy did not change. The relationship between the emotional mood of adolescents and beta frequency power during watching 3D films, with both passive and active glasses, was investigated in [26]. The results indicated high relative power of the beta band in passive glasses compared to the active ones. In [27], [28], areas of the brain that were sensitive to binocular vision and depth perception were shown. According to these studies, Brodmann areas of (BA) BA7 and BA19 of parietal and occipital cortex respectively had an important role in binocular stereo vision. BA37 and BA39 areas of the temporal cortex as well as the dorsal regions of occipital-temporal cortex [13] were sensitive to the power of depth perception. The area of the prefrontal cortex is depicted in Fig.1.

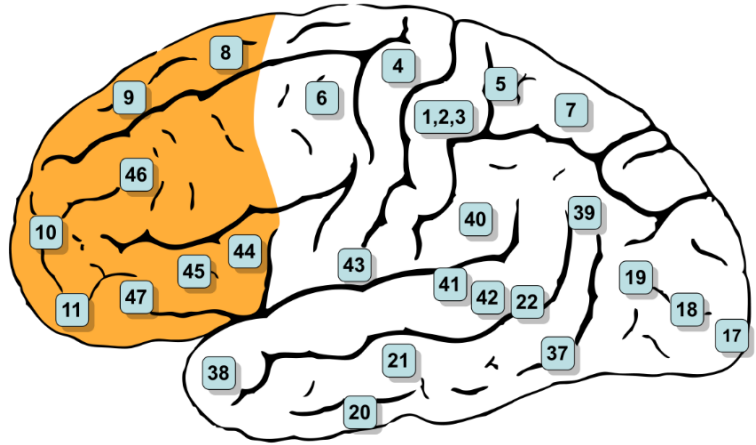

Fig. 1. Area of prefrontal cortex (This photo is taken from Wikipedia)

In this paper, the main purpose is to identify effective EEG frequency bands and, consequently, channels having a common feature between individuals as the result of watching 2D&3D video. As mentioned above, this research consists of a tripartite paradigm (a, b, c). In the first section, only the second section of test (b) is considered for 2D&3D analysis. In the second section, the experimental details are explained step by step. Spectrogram using short-time Fourier transform (STFT) is applied to EEG data in order to see the graphical understanding of EEG bands and their resolution in time segments. This visual representation offers the frequency spectrum of a signal that changes with time. Subsequently, the coefficients of STFT and, as a result, the power spectral density (PSD) are obtained for the EEG bands. In Section 3, the results based on time frequency analysis of the bands are presented. After a series of computational operations, the bands that are most affected by 2D variations over the 3D are selected as the dominant bands for the continuation of the test. Our results help us see the behavior of EEG channels in the selected bands. The discussion and conclusions are followed in Section 4.

## 2. Methods

The goal was to extract EEG signal features representing the highest percentage of PSD difference between 2D&3D. For this reason, Saw video [29] was selected to be shown to individuals in two different modes of 2D and 3D. Details of the test are described below.

### *2.1 Participants*

Eight subjects with the average age of 31±3.06 years old voluntarily participated in the test. Participants were doctoral students with normal visual acuity and do not have any medical history. Brief information of the test process was given to them before EEG recording. In order to reduce the effect of the artifact, participants were asked to keep unnecessary movements to a minimum. After the participants signed the informed consent form, the registration process was started. Karadeniz Technical University Medical Faculty in the date of 14.12.2018 has been decided by the Presidency of Scientific Research Ethics Committee that the study numbered 2018/286 is in accordance with the ethics committee's opinion. Ethical report number of the study was 24237859-806.

## 2.2. Experimental Setup

As explained above, the test consisted of three parts: (a, b, c). Signal recording included the following steps: A 9 sec. relaxation, watching 14 sec. of a 2D video and finally a 9 sec. resting. The duration of the movie was considered short due to the visual and mental fatigue of the participants [24]. To yield the EEG signals and fixing the signal-to-noise ratio [30], [31] each experiment was repeated 15 times. A few minutes after the 2D signal recording, exactly the same steps were repeated for the 3D. In this research, only video watching was reviewed. Passive 3D TV glasses were selected according to the LG 32 inch smart TV technology. Based on video display terminals (VDTs) [32] and because of the importance of the distance between the participant and the television screen as considerable environmental conditions, this distance was considered 130 cm. The TV was placed on a desk with a height of 120 cm and the background distance from the screen to the wall was 10cm. The main specifications of the LG 3D smart TV are presented in Table 1. A sampling rate of 512 Hz was selected for EEG recording. After EEG signal recording by a 21-channel cap (Brain Quick EEG System (Micromed, Italy)), the data preprocessing started. These channels were Fp1, Fpz, Fp2, F3, F4, F7, F8, C3, C4, Fz, P3, P4, Pz, O1, O2, T3, T4, T5, T6, Oz and Cz as the reference. Electrode locations as international 10-20 system is shown in Fig. 2. All the channels of this cap based on the 10-20 system were taken into consideration in the analysis.

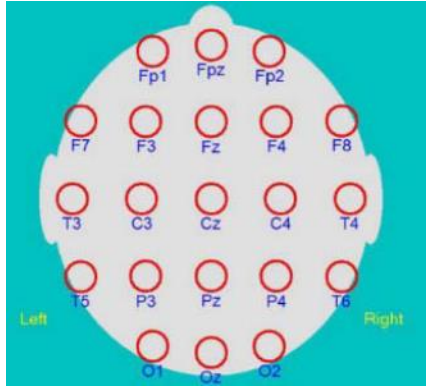

Fig. 2. Electrode locations as international 10-20 system

Table 1. The main specifications of the LG 3D smart TV [21]

| Variables | Model | Height | Refresh rate | Aspect ratio | Resolution | Size |
|---|---|---|---|---|---|---|
| Specifications | 32LN57 | 17.2 inch | 100Hz | 16:9 | $1920 \times 1080$ | 32 inch |

## 2.2.1 Basics of 3D TV

Choosing the type of glasses has a direct relationship with knowing how it works, and the full understanding of this issue requires the perception of the 3D basis. The human brain understands the surroundings in 3D because the image is a synthesis of the result from two human eyes in two different places by the brain. Due to the different location of the eyes [33] in the head, no object is understood at the same distance from both eyes. So, the 3D space for the brain is defined. This space makes it possible to understand the difference in distance. Due to the structure of the human eye, 3D TV and movies aim to create 3D illusions by providing different images to each eye. This is done using polarized lenses. The light-coated TV screen is polarized in different ways. This simple system works by combining two images of the TV screen for each eye. Due to the interaction of the lens polarization and the TV, the suitable image is prepared for each eye. The glasses that were used in the test had light weight and acceptable quality; they were battery-free and cost-effective. The ease of use and lightweight of these glasses were the main reason for their preference. Compared to the active type, the intensity of the lighting and its compatibility with the system were better.

## 2.3 Pre-processing of EEG Data

About the signal recording the duration of each trial was considered 14 sec. and in order to noise reduction, the average of 15 trials was analyzed for each channel [31]. The average of 15 trials was analyzed for each channel by applying 'mean' command of Matlab R2018b version. Then, a notch filter at 50 Hz was applied to suppress the line noise. A third order zero-phase bandpass Butterworth filter in the frequency range of 1-55 Hz was selected to clean up the noise signal. The non-linear phase effects were minimized in this way. The filter order was experimentally selected 3.

*2.4 Time Frequency Analysis and PSD Calculation and Band Selection*

Due to some disadvantages of EEG signals, e.g. small amplitude, being non-linear and non-stationary, and having a low noise-to-signal ratio, it is important to describe the frequency-time analysis [34]. In EEG studying, extracting the features that are accompanied by the reflection of the time, frequency and spatial characteristics are valuable. To demonstrate non-stationary EEG power distribution, PSD based on STFT is used. This transformation shows the distribution of power in the frequency-time range using the window function. There are several windows for calculating the spectral power density, but due to the unpredictable nature of the brain signals, a window with a smoothing behavior is selected. In this paper, the Hanning window function was chosen. In order to have proper time and frequency resolution, depending on the frequency range of the signal, the length of the window was selected as 512 sample length. The overlapping of the window was considered 'windowsize -1'. Now, to determine the difference between 2D&3D brain signals, the following method was considered. PSD of the EEG signal was calculated; $P_T$ represents total power density in the frequency range of 1 to 49 Hz. This variable was achieved by calculating the area under the curve of the PSD. The MATLAB '*trapz*' command was used for this purpose. $P_\delta$, $P_\theta$, $P_\alpha$, $P_\beta$ and $P_\gamma$ are the power density at 1-4, 4-8, 8-13, 13-25 and 25-49 Hz, respectively. For each band, the ratio of power to total power was computed. Then, the percentage of normalized power was taken into consideration. These steps were calculated individually for 2D&3D. For the 20 channels, the result was a matrix with the dimensions of 20 * 5 for each mode. The minus of values (2D-3D) in the matrices of 2D&3D made the brain behavior in movie video watching. These steps are shown in the band selection flowchart in Fig. 3. Through these steps and by comparing the results, the dominant bands were determined. Selection of these bands is presented in the results section. Accordingly, after calculating the normalized power differences in 2D and 3D, depending on the sign, the value of a difference greater than 2 would be meaningful for our study. After reviewing the results of all the participants, the band of channels meeting this requirement would be selected. Gamma band was not taken into account in other sections as it showed the 2D and 3D power spectrum difference in very few channels. Considering the results, the δ and θ bands were selected as dominant bands.

As mentioned above, in selecting the dominant bands, the duration of each trial was considered 14 sec. and the average of 15 trials was analyzed for each channel. Now, in the next section, we concentrat on EEG epoching to investigate specific time-windows extracting in order to check the behavior of each channel in trials.

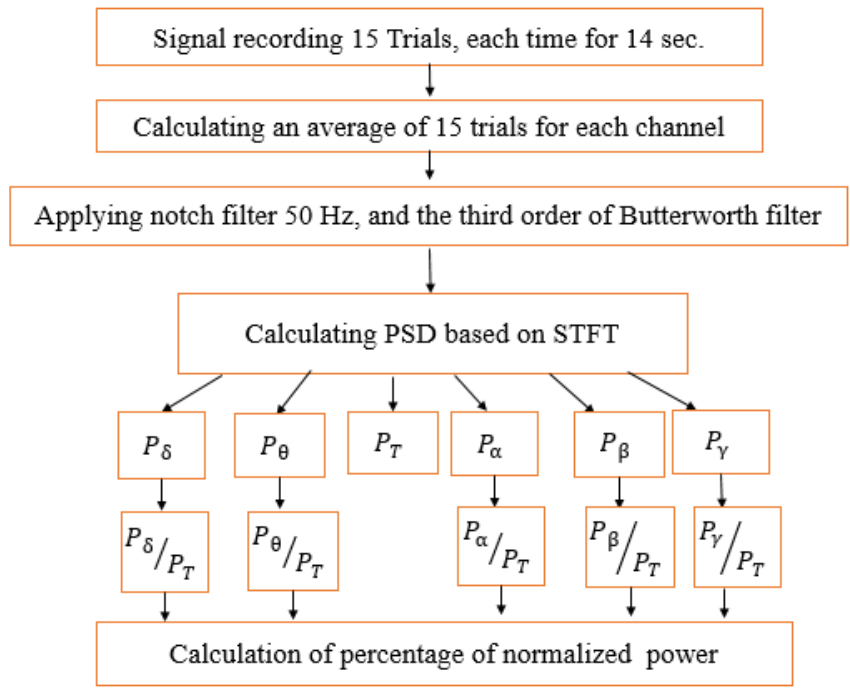

Fig. 3: Flowchart of band selection

*2.5 Epoch Category and Feature Extraction*

By applying a procedure to continuous EEG signals in order to obtain specific time-windows, epochs were extracted. As mentioned above, watching 2D or 3D movies for 14 sec. was repeated 15 times. The paradigm of this test is depicted below for better understanding.

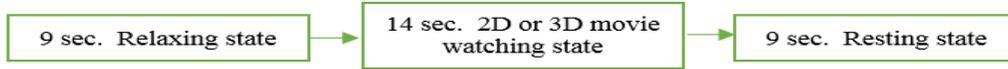

Fig. 4: Paradigm of EEG recording

After the signal recording, 4 sec. time segments with 3.5 sec. overlapping were used in this epoch category. Then, a notch filter at 50 Hz was applied to suppress the line noise. A third order zero-phase bandpass Butterworth filter in the frequency range of 1-35 Hz was selected to clean up the noise signal. The EEG frequency bands range was selected as (δ, 1-4), (θ, 4-8), (α, 8-12), (β, 12-30) after gamma deletion. $P_T$ represents total power density in the frequency range of 1 to 30 Hz. To extract the features, PSD based on STFT and relying on the two dominant bands, i.e. δ and θ, was considered. Dimensions of the feature extraction matrices for 2D or 3D would be 2*21*15. For each participant, there were 630 ½-sec. epochs in whole. There would be two classes in terms of 2D and 3D and 315 epochs per class. In order to prepare a dataset for classification, separating epochs of each class into two groups were performed. Now, training and testing sets were prepared. 158 and 157 epochs were chosen for training and testing sets, respectively. Flowchart of epoch category and feature extraction is depicted in Fig.5.

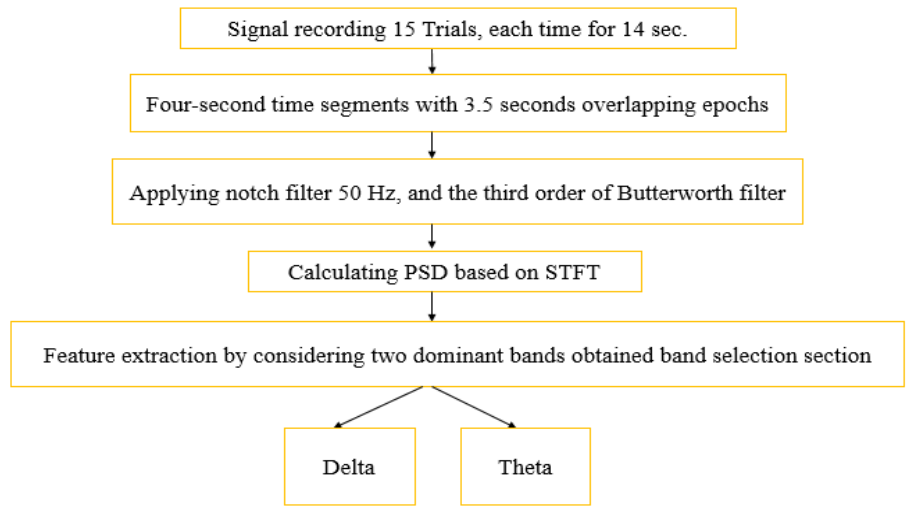

Fig. 5: Flowchart of epoch category and feature extraction

*2.6 Classification*

Due to the static complexity of the classification process, establishing a balance between time and frequency resolution is important. By following this, the extracted features are more effective and increase the recognition score. As we know, classification efficiency increases with the effectiveness of features. In this study, for the realization of the 2D&3D pattern classification, partial least squares regression (PLSR) and support vector machine (SVM) algorithms were employed. The basic definitions of these classifiers are given below.

### 2.6.1 PLSR Algorithm Using the Beta Distribution

Because of high computational and statistical efficiency, PLSR is selected by many researchers in different fields. This algorithm with excellent flexibility in terms of problem solving has the ability to reduce the high-dimensional datasets. In addition, PLSR is an algorithm that is combined with a regression model and can be used in different types of datasets of any dimensions. Another important feature of PLSR is its very high calculation speed; thus, this algorithm is believed to be very useful in diagnosing some health fields, e.g. tumour diagnosis. The mathematical expressions of this algorithm were described in detail in [35].

### 2.6.2 SVM Algorithm

Due to the unpredictable nature of the EEG signals, SVM can be a good candidate to separate the data via a hyper-plan with maximal margins [36]. As we know, each classification divides the dataset into the training and testing sets. Each sample of training set involves a class label and some different attributes (i.e. features). By designing a model that is based on the training data, SVM helps predict the class labels of the testing data. For this purpose, SVM algorithm is used kernel trick. Different kernel functions can be divided into four categories as linear, polynomial, radial basis function (RBF) and sigmoid [37]. To get satisfactory classification results, proper selection of parameters is important. RBF was used in this study and the related kernel function is defined in (1).

$$K(x,y) = e^{\frac{-1}{2}(\frac{|x-y|}{\sigma})^2} \tag{1}$$

In this equation, $\sigma$ is standard deviation of samples, and $x$ is the feature vector.

### 2.7 Cross-Validation

Applying cross-validation analysis to a new dataset in order to validate the results of classification algorithms is indispensable. This useful technique has the ability to evaluate predictive models. In other words, in classifying problems by using this technique, model performance will be tested on independent data (generally known as testing data). Reducing the runtime and increasing the accuracy of the analysis are the benefits of this approach.

*2.7.1 K-fold Cross-Validation*

In machine learning fields, *K*-fold cross-validation is a widespread type of cross-validation. The whole dataset is randomly divided into equal *K* subsamples. One of these subsamples holds as a testing set and the rest of *K*-1 data are in the training role. This process is repeated k folds. Then, in order to estimate the accuracy of the machine learning model, the results of all the rounds are averaged. In this study, the value of *K* is taken to be 10. After performing this cross-validation type, the β value of PLSR and the σ value of SVM are obtained.

*2.8 Evaluating the Performance of Classification*

As is known, one of the criteria for evaluating classifier performance is the classification accuracy (CA), which was presented as a percentage for the PLSR and SVM classifier in this study. This parameter is defined as the ratio of the number of correct predictions to the total number of predictions. In our study which had two classes, this parameter can be presented as follows:

$$CA = \frac{TP+TN}{TP+TN+FP+FN} \quad (2)$$

Where *TP* = True positives, *TN* = True negatives, *FP* = False positives, and *FN* = False negatives.

The classifier accuracy for analysing the performance of classification alone will not be sufficient in datasets that have a particularly unbalanced class. So, sensitivity and specificity are selected as the other criteria for determining the accuracy of the classification. In our study, we defined the 2D class as the positive samples and the 3D class as the negative samples; the sensitivity and specificity are shown below:

$$Sensitivity = \frac{TP}{TP+FN} \quad (3)$$

$$\text{Specificity} = \frac{TN}{TN+FP} \qquad (4)$$

The sensitivity represents a proportion of the true positive that has been correctly identified. In the same way, specificity indicates the proportion of actual negatives that has been correctly recognized.

## 3. Results

Among the five EEG bands, in 2D&3D video watching, PSD differences were observed in more channels in $\delta$ and $\theta$ bands. Therefore, these bands were chosen as the dominant bands in this study. For each channel, the average of eight participants' PSD differences of EEG bands is shown in Fig. 6. In Fig. 6, in the frontal region of the brain for $\delta$ band, the PSD of the 3D video movie viewers in comparison with 2D had a positive value i.e. (PSD (2D)-PSD (3D) < 0); but in parietal, central, temporal and occipital lobes of the brain, PSD of 2D viewers compared to the 3D one had a positive value i.e. (PSD (2D)-PSD (3D) > 0). At T6, for 2D&3D video viewers, the maximum difference of the average PSD was observed. For the $\theta$ band, among 20 channels in 2D&3D video watching, Oz had the maximum PSD difference. At Oz, PSD of the 3D viewer was more positive than the 2D one.

In order to classify the EEG signal powers of 2D&3D video watching, the behaviour of each channel was examined individually for eight participants. It was performed by using PLSR and SVM classifiers. The aim of this section is to determine the effective channels by looking at the average classification results of eight participants. Therefore, the average of the results of participants for both classifiers was calculated and reflected in Fig. 7. Considering this graph and its table below, it can be observed that the SVM classifier gave better results than PLSR. Since this study consisted of two classes, i.e. 2D&3D, the channels with percentage greater than 60 were chosen as the first stage of the channel selection. So, the best five channels for PLSR were as follows: O2, T4, Oz, T3 and Fz. In the same way, for the SVM classify the best channels are: O2, Oz, T4, T3, T5, F8, Fz, T6, C4, O1 and Fp1. The aim of ranking the best channels was to increase the percentage of classifier success. The best channel and their combinations would be tested as the second step in channel selection. In other words, for each classification method, both PLSR and SVM algorithms for these combinations would be tested. In the last step of the channel selection, it is important to pay attention to the compromise between the high success percentage and fewer number of channels. These channels and their combinations were tested with PLSR and SVM classifiers and are shown in Fig. 8 and Fig. 9. As seen in the figures, based on the above results, as expected, a striking increase in classification results was observed by applying this method. Based on the PLSR classifier, five channels were chosen as the best channel as mentioned above and the classification percentage of success was observed up to %75 and 89% increase for PLSR and SVM, respectively. For the SVM classifier, by keeping fourteen channels in mind, this result was increased to %81 and 97% for PLSR and SVM, respectively. In the SVM classifier, despite the good result, the great number of channels was not our desire; so we selected the classification result for the combination of eight channels with the success percentage of 79% and 93% for PLSR and SVM, respectively.

Beside the classification accuracy, the average sensitivity and specificity of all the participants for SVM classifier in the best channels section are shown in Fig. 10.

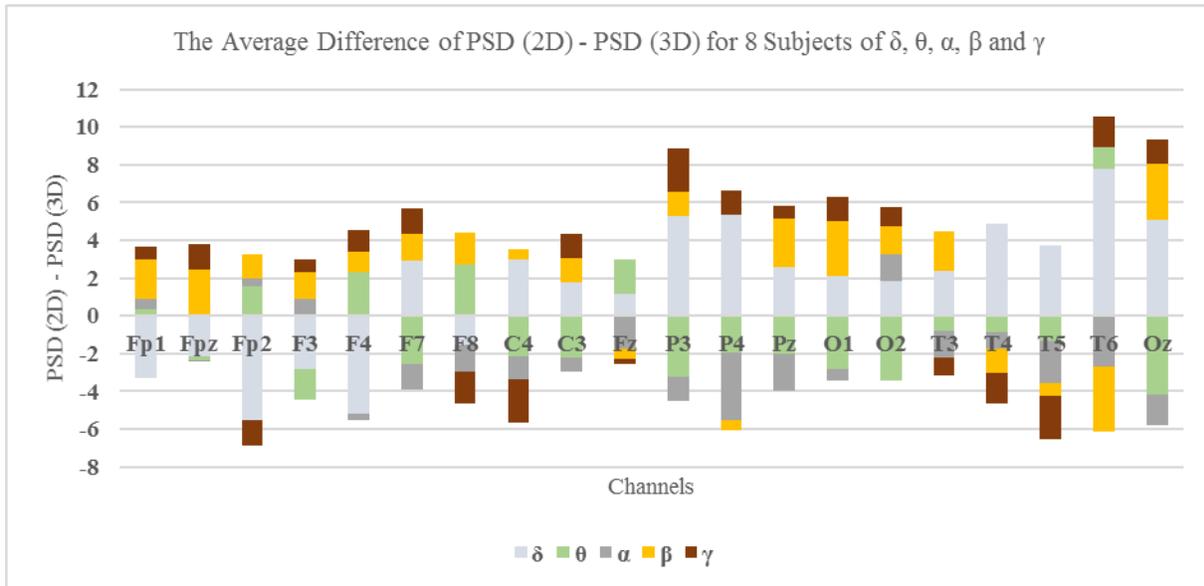

Fig. 6. The Average Difference of PSD (2D) - PSD (3D) for 8 Subjects of δ, θ, α, β and γ

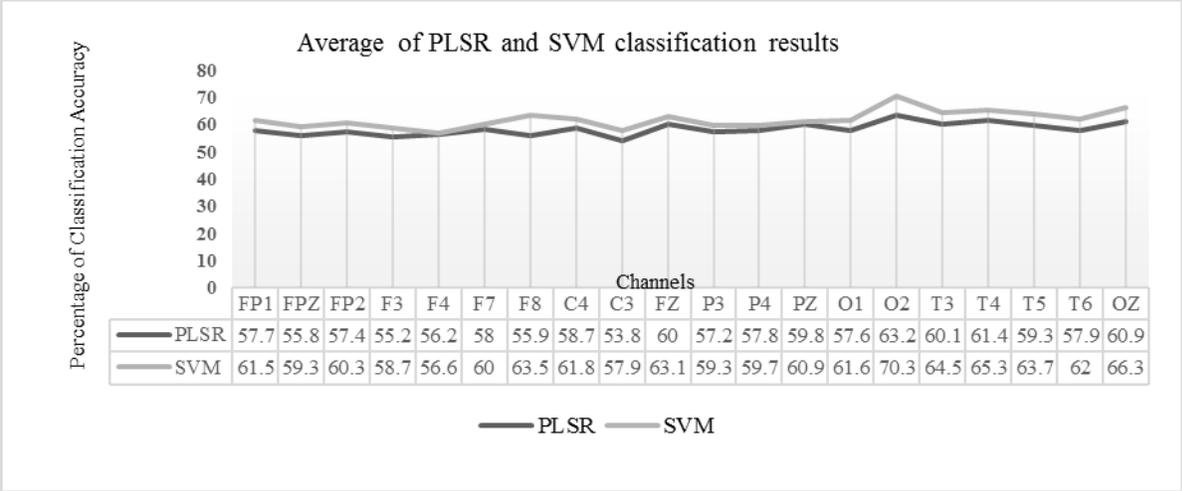

Fig. 7. Average of PLSR and SVM classification results

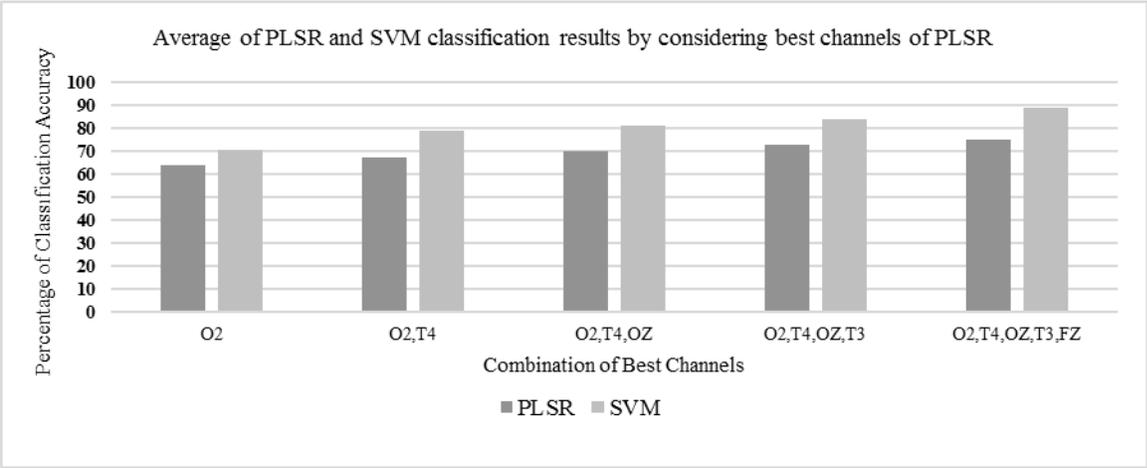

Fig. 8. Average of PLSR and SVM classification results by considering best channels of PLSR

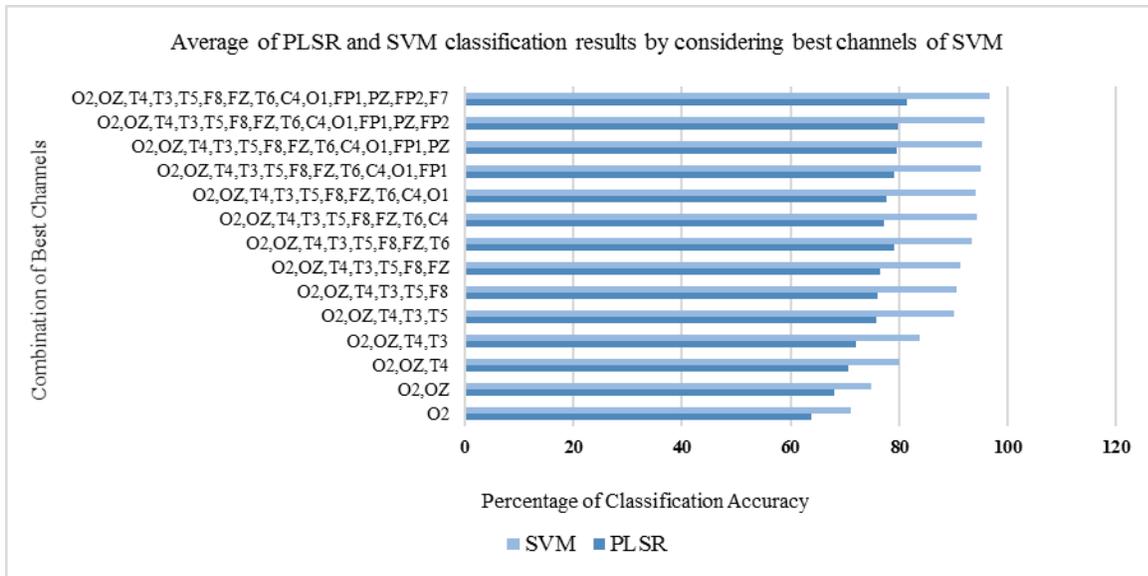

Fig. 9. Average of PLSR and SVM classification results by considering best channels of SVM

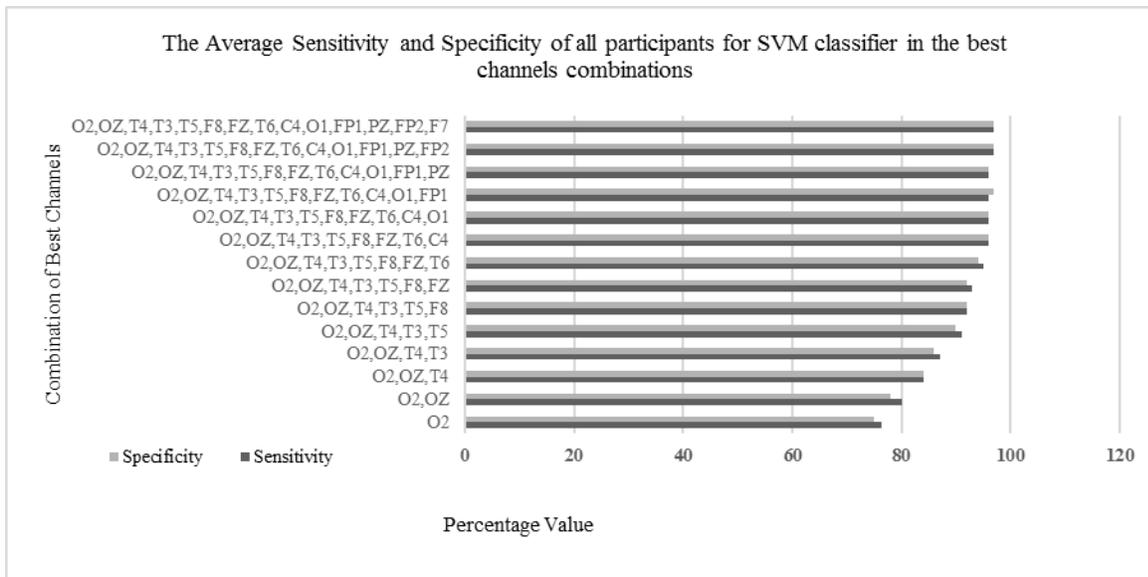

Fig. 10. Average sensitivity and specificity of all participants for SVM classifier in the best channels combinations

*4. Discussion*

In this study, all EEG bands were tested using the PSD of brain signals of 2D&3D video viewers. Then, δ and θ bands were used as the dominant bands to determine these modes. In [17], regardless of the behavior of the other bands, α and γ were considered for analysis based on the visual perception of the human brain. In 2D and 3D studies, there is no clear decision about the behavior of the brain waves in EEG bands. In this study, for each channel of each band considering the 2D&3D average PSD difference of all the participants, we selected the dominate bands which included more channels with maximum PSD difference.

The average PLSR and SVM classification results for each channel made the key channels of this research more significant. It is worth mentioning that the channels representing regions that were sensitive to stereo vision and depth perception [13] were included in our selected channels that gave the best results in this study. Channels starting with the letter O were those belonging to the occipital region and were responsible for the human vision. As expected in this study, these channels were included in the best channel combinations and, for both classification algorithms, they showed their roles. In this combination, the next important channels belonged to the temporal region. Among the different responsibilities of this lobe was the visual memory. The presence of channels from the frontal lobe can be described in terms of the importance of these channels in the focus of attention. The function of the parietal lobe is to interpret sensory information from different parts of the brain.

In order to evaluate the success of the classification performance, two criteria of sensitivity and specificity were used beside the classification accuracy. High values of these three factors showed a resultant binary classification. In our previous study [38], by using twelve electrode locations of the cap, for all the EEG bands, the value of these factors was lower than this proposed research. These factors increased up to an acceptable value thanks to the correct selection of bands and channels.

*5. Conclusion*

In this paper, we investigated the effects of 2D&3D video effects on the human brain signals. The PSD of the EEG recording of 2D and 3D film viewers was chosen as the main parameter of this study. In 2D&3D video watching and EEG recording studies, the principle of using the candidate channels representing all the brain regions and testing the effects of all EEG bands could be considered the privilege of this study. There are very few studies that investigate these effects in detail. In this work, after testing all the EEG bands, the average power spectrum difference of eight adult 2D and 3D viewers was observed to be significant in the δ and θ bands for most channels. In the feature extraction step, after separating the EEG signals into certain epochs, a PSD based on STFT was used and two dominant bands were extracted for each channel as a feature. As known, more expressive features enhance classification performance; hence, feature extraction has become the most critically significant step in EEG data classification. In the channel selection section, PLSR and SVM classification algorithms were selected to separate the EEG signals of 2D and 3D video film viewers. Acceptable results were obtained for PLSR and SVM algorithms after selecting the effective channels. Admittedly, the findings of this study helped clarify the human brain affected by 2D&3D videos, as well as the temporary properties of the EEG bands and channels.

In the future study, the number of participants could be increased to solve one of the limitations of this study. In this work which consisted of a 3-stage paradigm, it was aimed to analyze the moment of transition from watching the video to relaxation and from watching to rest. We need to emphasize that this is initial research and a preliminary study of our hypothesis. As we know, people see their surroundings as 3D because of their eye structures. When people fall asleep, we can claim that they will lose their perception of depth, in which case there may be a sudden transition from 3D to 2D. Capturing this important moment is our main goal. In the power spectrum analysis of the 2D and 3D EEG signals, we tried to classify these signals with the highest percentage of success by using channels and dominant bands representing the lobes of the brain. Reducing the number of channels and increasing the classification accuracy by using different feature extraction and classification methods could be targeted in our future study.

## *References*

1. Voiculescu, M., Segarceanu, A., Negutu, M., Ghita, I., Fulga, I., & Oa, C. (2015). The effect of caffeine on cerebral asymmetry in rats. *Journal of medicine and life*, *8*(4), 476–482.
2. Ahmed, S. M., & Abbas, S. N. (2015). A New EEG Acquisition Protocol for Biometric Identification Using Eye Blinking Signals. *International Journal of Intelligent Systems and Applications*, *7*(6,May), 48. doi:10.5815/ijisa.2015.06.05
3. Lianyang Li, Pagnotta, M. F., Arakaki, X., Tran, T., Strickland, D., Harrington, M., … Subjects, A. (2015). Brain

Activation Profiles in mTBI : Evidence from Combined Resting-State EEG and MEG Activity. *2015 37th Annual International Conference of the IEEE Engineering in Medicine and Biology Society (EMBC)*, 6963–6966. doi:10.1109/EMBC.2015.7319994

4. Fraga, F. J., Quispe, G., Johns, E., Tavares, G., Falk, T. H., & Phillips, N. A. (2018). Computer Methods and Programs in Biomedicine Early diagnosis of mild cognitive impairment and Alzheimer ' s with event-related potentials and event-related desynchronization in N-back working memory tasks. *Computer Methods and Programs in Biomedicine*, *164*, 1–13. doi:10.1016/j.cmpb.2018.06.011

5. Tzallas, A. T., Tsipouras, M. G., Fotiadis, D. I., & Member, S. (2009). Epileptic Seizure Detection in EEGs Using Time – Frequency Analysis. *IEEE transactions on information technology in biomedicine*, *13*(5), 703–710.

6. Kober, S. E., Schweiger, D., Witte, M., Reichert, J. L., Grieshofer, P., Neuper, C., & Wood, G. (2015). Specific effects of EEG based neurofeedback training on memory functions in post-stroke victims. *Journal of NeuroEngineering and Rehabilitation*, 1–13. doi:10.1186/s12984-015-0105-6

7. Angela, R., Billeci, L., Crifaci, G., Troise, E., Tortorella, G., & Pioggia, G. (2016). Research in Developmental Disabilities Cognitive training modifies frequency EEG bands and neuropsychological measures in Rett syndrome. *Research in Developmental Disabilities*, *53–54*, 73–85. doi:10.1016/j.ridd.2016.01.009

8. Trudeau, D. L. (2008). EEG Biofeedback as a Treatment for Substance Use Disorders : Review , Rating of Efficacy , and Recommendations for Further. *Journal of Neurotherapy*, *12*(1), 1–28. doi:10.1007/s10484-007-9047-5

9. Kang, J., Chang, Y., & Kim, S. (2018). Neurocomputing Electroencephalographic feature evaluation for improving personal authentication performance. *Neurocomputing*, *287*, 93–101. doi:10.1016/j.neucom.2018.01.074

10. NeurotechEDU. (n.d.). Retrieved March 6, 2019, from http://learn.neurotechedu.com/

11. EEG - ECG - Biosensors. (n.d.). Retrieved March 6, 2019, from http://neurosky.com/

12. Lenc, T., Keller, P. E., Varlet, M., & Nozaradan, S. (2018). Neural tracking of the musical beat is enhanced by low-frequency sounds. *Proceedings of the National Academy of Sciences*, *115*(32), 8221–8226. doi:10.1073/pnas.1801421115

13. Zwezdochkina, N., & Antipov, V. (2018). The EEG Activity during Binocular Depth Perception of 2D Images. *Computational intelligence and neuroscience*, *2018*, 1–7. doi:10.1155/2018/5623165

14. Mumtaz, W., Xia, L., Malik, A. S., Member, S., Azhar, M., & Yasin, M. (2013). EEG Classification of Physiological Conditions in 2D / 3D Environments Using Neural Network *. *2013 35th Annual International Conference of the IEEE Engineering in Medicine and Biology Society (EMBC)*, *1*(3), 4235–4238. doi:10.1109/EMBC.2013.6610480

15. Khairuddin, H. R., Malik, A. S., Member, S., Mumtaz, W., Kamel, N., Member, S., & Member, L. X. (2013). Analysis of EEG Signals Regularity in Adults during Video Game Play in 2D and 3D. *2013 35th Annual International*

*Conference of the IEEE Engineering in Medicine and Biology Society (EMBC)*, 2064–2067. doi:10.1109/EMBC.2013.6609938

16. Khairuddin, R. N. H. R., Malik, A. S., & Kamel, N. (2014). EEG Topographical Maps Analysis for 2D and 3D Video Game Play. *2014 5th International Conference on Intelligent and Advanced Systems (ICIAS)*, (June), 1–4. doi:10.1109/ICIAS.2014.6869517

17. Kim, S., & Kim, D. (n.d.). Differences in the Brain Waves of 3D and 2 . 5D Motion Picture Viewers. *arXiv preprint arXiv:1210.2147*.

18. Han, Y., Lin, H. Y., & Chen, C. (2017). SP-3 Visual Fatigue for Laser-Projection Light-Field 3D Display in Contrast with 2D Display. *2017 24th International Workshop on Active-Matrix Flatpanel Displays and Devices (AM-FPD*, 9–12.

19. Bamatraf, S., Hussain, M., Aboalsamh, H., Mathkour, H., Malik, A. S., Amin, H. U., … Qazi, E. (2015). A System based on 3D and 2D Educational Contents for True and False Memory Prediction using EEG Signals. *2015 7th International IEEE/EMBS Conference on Neural Engineering (NER)*, 1096–1099.

20. Avarvand, Forooz Shahbazi and Bosse, Sebastian and Muller, Klaus-Robert and Schufer, Ralf and Nolte, Guido and Wiegand, Thomas and Curio, Gabriel and Samek, W. (2017). Objective quality assessment of stereoscopic images with vertical disparity using EEG. *Journal of neural engineering*, *14*(4), 1–14.

21. Ramadan, M. Z., Alhaag, M. H., Abidi, M. H., Ramadan, M. Z., Alhaag, M. H., Abidi, M. H., & Haider Abidi, M. (2017). Effects of Viewing Displays from Different Distances on Human Visual System. *Applied Sciences*, *7*(11), 1153. doi:10.3390/app7111153

22. Manshouri, N., Maleki, M., & Kayıkçıoğlu, T. (2017). Classification of Human Vision Discrepancy during Watching 2D and 3D Movies Based on EEG Signals. *International Journal of Computer Science and Information Security*, *15*(2), 430–436.

23. Kim, Y.-J., & Lee, E. C. (2011). EEG Based Comparative Measurement of Visual Fatigue Caused by 2D and 3D Displays (pp. 289–292). Springer, Berlin, Heidelberg. doi:10.1007/978-3-642-22095-1_59

24. Subasi, A. (2005). Automatic recognition of alertness level from EEG by using neural network and wavelet coefficients. *Expert systems with applications*, *28*(4), 701–711. doi:10.1016/j.eswa.2004.12.027

25. Chen, C., Li, K., Wu, Q., Wang, H., Qian, Z., & Sudlow, G. (2013). EEG-based detection and evaluation of fatigue caused by watching 3DTV. *Displays*, *34*(2), 81–88. doi:10.1016/j.displa.2013.01.002

26. Park, S. J., Subramaniyam, M., Moon, M. K., & Kim, D. G. (2013). Physiological Responses to Watching 3D on Television with Active and Passive Glasses Physiological Responses to Watching 3D on Television. *International Conference on Human-Computer Interaction*, (January 2018), 498–502. doi:10.1007/978-3-642-39473-7

27. Fischmeister, F. P. S., & Bauer, H. (2006). Neural correlates of monocular and binocular depth cues based on natural

images : A LORETA analysis. *Vision research*, *46*(20), 3373–3380. doi:10.1016/j.visres.2006.04.026

28. Rutschmann, R. M., & Greenlee, M. W. (2004). N EURO R EPORT BOLD response in dorsal areas varies with relative disparity level. *Neuroreport*, *15*(4), 615–619. doi:10.1097/01.wnr.0000094160.86963.8c

29. *(1) 3D Video Chain Saw! - YouTube*. (n.d.). Retrieved from https://www.youtube.com/watch?v=foQNrtUsEjw

30. Smith, S. J. M. (2005). EEG in the diagnosis, classification, and management of patients with epilepsy. *Journal of neurology, neurosurgery, and psychiatry*, *76 Suppl 2*(suppl 2), ii2-7. doi:10.1136/jnnp.2005.069245

31. Repovš, G. (2010). *Repovš G: Dealing with Noise in EEG Recording and Data Analysis 18 Dealing with Noise in EEG Recording and Data Analysis Spoprijemanje s šumom pri zajemanju in analizi EEG signala*. *Informatica Medica Slovenica* (Vol. 15). Retrieved from http://ims.mf.uni-lj.si/archive/15(1)/21.pdf

32. Patterson, R. (2007). Human factors of 3-D displays. *Journal of the Society for Information Display*, *15*(11), 861. doi:10.1889/1.2812986

33. Forrester, J. V., Dick, A. D., McMenamin, P. G., Roberts, F., Pearlman, E., Forrester, J. V., … Pearlman, E. (2016). Physiology of vision and the visual system. *The Eye*, 269–337.e2. doi:10.1016/B978-0-7020-5554-6.00005-8

34. Kim, C., Sun, J., Liu, D., Wang, Q., & Paek, S. (2018). An effective feature extraction method by power spectral density of EEG signal for 2-class motor imagery-based BCI. *Medical \& biological engineering \& computing*, *56*(9), 1645–1658.

35. Boulesteix, A., & Strimmer, K. (2006). Partial least squares : a versatile tool for the analysis of high-dimensional genomic data. *Briefings in bioinformatics*, *8*(1), 32–44. doi:10.1093/bib/bbl016

36. Amin, H. U., Mumtaz, W., & Subhani, A. R. (2017). Classification of EEG Signals Based on Pattern Recognition Approach. *Frontiers in computational neuroscience*, *11*(November), 1–12. doi:10.3389/fncom.2017.00103

37. Goel, A. (2016). Role of Kernel Parameters in Performance Evaluation of SVM. *2016 Second International Conference on Computational Intelligence & Communication Technology (CICT)*, 166–169. doi:10.1109/CICT.2016.40

38. Manshouri, N., & Kayikcioglu, T. (2016). Classification of 2D and 3D Videos Based on EEG Waves. *2016 24th Signal Processing and Communication Application Conference (SIU)*, 949–952. doi:10.1109/SIU.2016.7495898